# Generalized "Square roots of Not" matrices, their application to the unveiling of hidden logical operators and to the definition of fully matrix circular Euler functions

**Eduardo Mizraji**

Group of Cognitive Systems Modeling, Biophysics and Systems Biology Section,
Facultad de Ciencias, Universidad de la República
Iguá 4225, Montevideo 11400, Uruguay
Emails: emizraji@gmail.com, mizraj@fcien.edu.uy
ORCID ID: 0000-0001-6938-8427

**ABSTRACT**

The square root of Not is a logical operator of importance in quantum computing theory and of interest as a mathematical object in its own right. In physics, it is a square complex matrix of dimension 2. In the present work it is a complex square matrix of arbitrary dimension. The introduction of linear algebra into logical theory has been enhanced in recent decades by the researches in the field of neural networks and quantum computing. Here we will make a brief description of the representation of logical operations through matrices and we show how general expressions for the two square roots of the Not operator are obtained. Then, we explore two topics. First, we study an extension to a non-quantum domain of a short form of Deutsch's algorithm. Then, we assume that a square root of Not is a matrix extension of the imaginary unit i, and under this idea we obtain fully matrix versions for the Euler representations of circular functions by complex exponentials.

## 1. Introduction.

The remarkable properties of the square roots of Not, which are complex square matrices, emerged from research on quantum computing. This theory is currently producing great interest in the representation of logical operations in terms of linear algebra. In fact, at least two areas of research have led in recent decades to represent basic logical functions using vectors and matrices, on the one hand the theory of quantum computing [2, 6] and on the other, research on neural models of associative memories and reasoning [16, 17].

In the theory of quantum computing, where the square root of not (SRN) operator was born, logical variables are represented by two-dimensional vectors called "qubits" that represent quantum states [2]. This causes that in this framework some logical operators, in particular the SRNs, are square matrices of dimension 2. The mathematical structure of the SRN in the quantum context and its remarkable physical properties were investigated in [6].

In this work we will use a representation that has been called "Vector logic" [14, 15] to analyze various properties of the SRN, both as a logical operator and as an algebraic object. The formalism of vector logic had its origin in neuroscience, especially in the representation of associative memories through large-dimension matrices. This formalism allows the SRN research to be extended in an interesting way, both in its structure, which is generalized to large-dimensional vectors, and in its applications to various problems. Recently the generalized matrix version of the SRN was used to represent the virtuality of counterfactual propositions [16].

To prepare the study of the SRN, we will first show the basic logical functions using truth tables, then an arithmetic counterpart of those functions and we will finish the presentation of the basic data by showing a representation of those logical functions using vectors and matrices. Then we will show an elementary deduction of the very aesthetic structure of the two SRNs for vectors of arbitrary dimension. Finally, we show two heuristic explorations. One concerns the fascinating fact that SRN allows us to capture information about the structure of hidden logic gates using a single minimal input. The other is on the search for purely matrix representations of the beautiful Eulerian expressions of circular functions, substituting the imaginary unit $\mathrm{i} = \sqrt{-1}$ for an SRN.

## 2. The road from classic logic to matrix algebra.

Plausibly the link between classical logic and mathematics began with Leibniz and then took a strong impetus in the 19th century with the Cambridge algebraic movement. This movement sought to incorporate Leibniz's symbolic formalisms for differential and integral calculus in British mathematics (for a history of this process see Diagne [7] and Houser [8]). It is in this context that George Boole, whose primary specialty was the analysis of differential equations using operational procedures, becomes involved in logic research. This framework of search for new formalisms led Boole to publish his seminal work on logic, "An Investigation of the Laws of thought" [3]. These investigations led to the possibility of expressing logical operations by means of algebraic equations defined on parameters that represented the truth values. The more classical representation assigns the number 1 to the value "true" and the number 0 to "false", but we will see later by means of an example that this is only one of many interesting representations. Cayley's founding work on matrix algebra [4] was published a few years before Boole's death. The initial link of logic with matrix theory was proposed by Charles Peirce and then formally established in Irving Copi's 1948 article (signed Copilowich [5]).

In the field of neuroscience, the pioneering model that created the area of neurocomputing, published by McCulloch and Pitts in 1943, was completely based on the mathematical representation of logic [12]. This established a very early link between logic and computational neuroscience. With the development of the theory of matrix associative memories, particularly by Kohonen [10] the natural interest arose to analyze whether logical operations could be represented by memory modules. In [14] it is shown that this is possible and this formalism, born in the theory of neural networks, showed a great capacity to represent a variety of logical operations through linear algebra [14, 15]. It is a remarkable fact that the Polish notation developed by Lukasiewicz generates an ordering of logical functions and variables exactly the same as that obtained by representing logical operations by means of matrices and vectors [11]. This shows that, in fact, Jan Lukasiewicz was a forerunner in approaching logic from the perspective of a theory of operators.

In physics, very early, with the origin of quantum mechanics, a theory based on matrix and tensor formalisms was developed to analyze a novel domain, inspired by the complexities of the foundations of quantum physics. This domain of research was called "quantum logic" [13]. In addition, in recent years, a very vigorous effort has been developed in research on quantum computing, provoking a renewed and strong interest in the representation of logical operations in terms of matrices and vectors [2, 6].

**2.1. Basic logic functions.**

The formalization of classical logical functions begins by defining a pair of abstract objects, the truth values "true", t, and "false", f. These objects can take the structure that suits the situation (words, numbers, vectors, etc.). By noting the set of truth values as $\tau = \{t, f\}$ two classes of basic logical operations are defined:

(a) Monadic operations, represented by functions $\mathrm{Mon} : \tau \to \tau$ .

(b) Dyadic operations, represented by functions $\mathrm{Dyad} : \tau \times \tau \to \tau$ ($\times$ means Cartesian product). The table representing the four monadic functions, Identity Id, Negation Not, Constant affirmation Cid and Constant negation Cnot, is as follows.

TABLE 1

| *p* | *Id* | *Not* | *Cid* | *Cnot* |
|---|---|---|---|---|
| t | t | f | t | f |
| f | f | t | t | f |

The following table represents the dyadic functions Implication IMPL, Disjunction OR, Conjunction AND, Equivalence EQUI and Exclusive-or XOR:

TABLE 2

| *p* | *q* | *IMPL(p,q)* | *OR(p,q)* | *AND(p,q)* | *EQUI(p,q)* | *XOR(p,q)* |
|---|---|---|---|---|---|---|
| t | t | t | t | t | t | f |
| t | f | f | t | f | f | t |
| f | t | t | t | f | f | t |
| f | f | t | f | f | t | f |

**2.2. An arithmetic of logic functions based on 1 and -1.**

The usual way to arithmetic the operations of the logic shown in the previous tables is to assign the 1 to the value "true" and the 0 to "false". This leads to the classic Boolean representations. But it is interesting to study the arithmetic representation of these functions that is obtained if we establish the correspondences $t \to 1$ and $f \to -1$, with a truth values set $\tau = \{1, -1\}$.This representation is important in the context of this article, because it allows us to obtain the Not function under the form of a scalar operator, and from there to achieve a scalar version of the root of Not. Thus, it can be seen that the four monadic operations for are given by the following equations:

$$\mathrm{Id}(w) = 1.w \ , \ \mathrm{Not}(w) = -1.w \ , \ \mathrm{Cid}(w) = 1.w^2 \ , \ \mathrm{Cnot}(w) = -1.w^2 \ .$$

Note that both the identity and the negation can be symbolized in this arithmetic by independent numerical operators separable from the variables: $\mathrm{Id} = 1$ and $\mathrm{Not} = -1$. This is an interesting point for the arguments that we will see in future sections (and that is not achieved if the usual Boolean arithmetic values 1 and 0 are chosen).

The dyadic functions represented in the table are described for the $u, v \in \tau$ truth values by the following equations:

$$\mathrm{IMPL}(u,v) = \frac{-u+v}{2} + \left[1 - \left(\frac{-u+v}{2}\right)^2\right] uv \ ,$$

$$\mathrm{OR}(u,v) = \frac{u+v}{2} - \left[1 - \left(\frac{u+v}{2}\right)^2\right] uv \ ,$$

$$\mathrm{AND}(u,v) = \frac{u+v}{2} + \left[1 - \left(\frac{u+v}{2}\right)^2\right] uv,$$

$$\mathrm{EQUI}(u,v) = uv,$$

$$\mathrm{XOR}(u,v) = -uv \ .$$

It is interesting to note that, as is expected for any consistent representation, these equations satisfy the relationship between implication and disjunction:

$$\mathrm{IMPl}(u,v) = \mathrm{OR}[\mathrm{Not}(u), v] \qquad (1)$$

and also DeMorgan's law:

$$\mathrm{OR}(u,v) = \mathrm{Not}[\mathrm{AND}(\mathrm{Not}(u), \mathrm{Not}(v)] \ . \qquad (2)$$

**2.3 Vectors, matrices and logical operations.**

Now we will describe these logical functions representing the truth values by vectors. Let us take two column vectors $s, n \in \mathbb{R}^{Q\times 1}$ as truth values, and assign each vector to the symbolic truth values as follows: $t \to s$ and $f \to n$. Hence $\tau = \{s, n\}$. From now on we will assume that the vectors are normal. This means that both modules are $|s| = |n| = 1$. This is equivalent to put $s^T s = \langle s, s \rangle = 1$ and $n^T n = \langle n, n \rangle = 1$, where the superindex T means transposition and $\langle \ , \ \rangle$ represents the inner product.

In this framework, all monadic functions can be represented by square matrices $U \in \mathbb{R}^{Q\times Q}$ such that

$$Us = a, \ Un = b \ , \quad a, b \in \{s, n\} \ . \qquad (3)$$

To find the structure of this matrix, the following procedure can be established. Equations (3) imply this matrix equation (this procedure, used to solve these monadic operators, and then the dyadic matrices and the SRNs, was originally used by Kohonen [10] to establish the structure of associative matrix memories and then adapted to the logic problems in [14]:

$$U[s\,n] = [a\,b] \ . \qquad (4)$$

$[s\,n]$ y $[a\,b]$ are partitioned matrices of order. $Q \times 2$. This equation (4) has the formal solution

$$U = [a\,b][s\,n]^{+} \ , \qquad (5)$$

$[s\,n]^{+}$ being a pseudoinverse (see Barnett [1]). For linearly independent column s and n vectors, the pseudoinverse has an exact solution given by

$$[s\,n]^{+} = \left([s\,n]^{T}[s\,n]\right)^{-1}[s\,n]^{T} \quad . \tag{6}$$

Suppose that the orthonormal vectors s and n have the following scalar product: $\langle s,n\rangle = \langle n,s\rangle = \varepsilon$ with $\varepsilon \in (0.1)$. Under these conditions it is found that the explicit solution of equation (4) is

$$[s\,n]^{+} = \frac{1}{1-\varepsilon^{2}}\begin{bmatrix} s^{T} - \varepsilon n^{T} \\ n^{T} - \varepsilon s^{T} \end{bmatrix} .$$

Defining the pseudo-true vector y and the pseudo-false vector z as

$$y = [1/(1-\varepsilon^{2})]\left(s - \varepsilon n\right) \; ; \;\; z = [1/(1-\varepsilon^{2})]\left(n - \varepsilon s\right),$$

we get

$$[s\,n]^{+} = [y\,z]^{T} .$$

Consequently, the monadic matrices U have the following general structure:

$$U = [a\,b][y\,z]^{T} = ay^{T} + bz^{T} \quad . \tag{7}$$

Note that $\langle y,s\rangle = \langle z,n\rangle = 1$ , and $\langle y,n\rangle = \langle z,s\rangle = 0$. Hence, $U s = a$ and $U n = b$.

In this framework, if the truth values are orthogonal (y = s and z = n), the operators identity I, negation N, constant affirmation K and constant negation M, have the following structure:

$$I = ss^{T} + nn^{T} \;\; ; \;\; N = ns^{T} + sn^{T} ; \;\; K = ss^{T} + sn^{T} \;\; ; \;\; M = ns^{T} + nn^{T} .$$

For identity and negation it turns out that $I s = s$, $I n = n$ and $N s = n$ , $N n = s$. On the other hand, $K s = K n = s$ and $M s = M n = n$. Those results are the matrix version of those shown in Table 1.

The matrix representation of the dyadic functions is based on the properties of the Kronecker product [9]. Let us start by defining this product and showing the properties that interest us here. The Kronecker product, is usually defined as follows: Let U and V be two matrices. These matrices can be of any dimension, included vectors. The definition of the Kronecker product $U \otimes V$ is

$$U \otimes V = \left[ u_{ij} V \right] .$$

We now numerically exemplify this definition

$$U = \begin{bmatrix} 1 & 0 \\ 2 & -1 \end{bmatrix}_{2\times 2} \; ; \; V = \begin{bmatrix} 1 & -1 & 4 \\ 3 & 1 & 0 \end{bmatrix}_{2\times 3} \; ; \; U \otimes V = \begin{bmatrix} 1V & 0V \\ 2V & -1V \end{bmatrix} = \begin{bmatrix} 1 & -1 & 4 & 0 & 0 & 0 \\ 3 & 1 & 0 & 0 & 0 & 0 \\ 2 & -2 & 8 & -1 & 1 & -4 \\ 6 & 2 & 0 & -3 & -1 & 0 \end{bmatrix}_{4\times 6}$$

The two properties of this product relevant to the representation of dyadic logical operators are the following:

(a) $(U \otimes V)^T = U^T \otimes V^T$

(b) $(U \otimes V)(U' \otimes V') = (UU') \otimes (VV')$

This property (b) requires that each of the pairs U and U 'as well as V and V' be conformable for the product. Let us show these properties for four q-dimensional column vectors a, b, c, and d:

$$(a \otimes b)^T (c \otimes d) = (a^T \otimes b^T)(c \otimes d)$$

Therefore, by property (b),

$$(a \otimes b)^T (c \otimes d) = (a^T c)(b^T a) = \langle a,c \rangle \langle b,d \rangle . \qquad (8)$$

This is the cardinal property that allows the computation of dyadic operations through matrices [14].

Given the truth vectors s and n, these dyadic logic matrices T must have the following computational capabilities:

$$T(s \otimes s) = e\,;\; T(s \otimes n) = f\;\;;\;\; T(n \otimes s) = g\,;\; T(n \otimes n) = h\,,$$

for $e, f, g, h \in \{s, n\}$ .

Following a calculation procedure analogous to the one shown here for monadic functions, if the vectors s and n are linearly independent, the matrix T has an exact solution:

$$T = [e\ f\ g\ h]\left([s\,n]^{+} \otimes [s\,n]^{+}\right) = [e\ f\ g\ h]\left([y\,z]^{T} \otimes [y\,z]^{T}\right)\ .$$

Developing this equation, the result is the following:

$$T = e(y \otimes y)^{T} + f\,(y \otimes z)^{T} + g(z \otimes y)^{T} + h\,(z \otimes z)^{T}\ . \qquad (9)$$

Note that $T \in \mathbb{R}^{Q \times Q^2}$ . If the vectors are orthogonal, then y = s and z = n. For this situation the matrices implication L, disjunction D, conjunstion C, equivalence E and exclusive-or X, are the following:

$$L = s(s \otimes s)^{T} + n(s \otimes n)^{T} + s(n \otimes s)^{T} + s(n \otimes n)^{T}\ ,$$

$$D = s(s \otimes s)^{T} + s(s \otimes n)^{T} + s(n \otimes s)^{T} + n(n \otimes n)^{T},$$

$$C = s(s \otimes s)^{T} + n(s \otimes n)^{T} + n(n \otimes s)^{T} + n(n \otimes n)^{T},$$

$$E = s(s \otimes s)^{T} + n(s \otimes n)^{T} + n(n \otimes s)^{T} + s(n \otimes n)^{T},$$

$$X = n(s \otimes s)^{T} + s(s \otimes n)^{T} + s(n \otimes s)^{T} + n(n \otimes n)^{T}.$$

From equation (8), we can easily check that

$$L(s \otimes s) = s\;\;;\; L(s \otimes n) = n\;\;;\; L(n \otimes s) = s\,;\; L(n \otimes n) = s\,.$$

This matrix version of the implication corresponds to the definition shown in Table 2 for IMPL (p, q). Let us point out that these monadic and dyadic logical matrices give us a version in terms of operators of the important tautologies shown in equations (1) and (2):

$$L = D(N \otimes I) \ , \qquad (1bis)$$

$$D = NC(N \otimes N) \ . \qquad (2bis)$$

This matrix version of the logic is what will allow us to analyze the Square Roots of Not. To finish this matrix presentation of logic, we show the structure of some of the logical logic matrices for two 2D orthogonal bases:

Set 1: $s = \begin{bmatrix} 1 \\ 0 \end{bmatrix} \ , \ n = \begin{bmatrix} 0 \\ 1 \end{bmatrix}$ ; Set 2: $s = \frac{1}{\sqrt{2}}\begin{bmatrix} 1 \\ 1 \end{bmatrix} \ , \ n = \frac{1}{\sqrt{2}}\begin{bmatrix} 1 \\ -1 \end{bmatrix}$.

*Case 1*. Identity and negation operators for Set 1 and Set 2.

Set 1: $I = \begin{bmatrix} 1 & 0 \\ 0 & 1 \end{bmatrix}, \ N = \begin{bmatrix} 0 & 1 \\ 1 & 0 \end{bmatrix},$

Set 2: $I = \begin{bmatrix} 1 & 0 \\ 0 & 1 \end{bmatrix}, \ N = \begin{bmatrix} 1 & 0 \\ 0 & -1 \end{bmatrix},$

*Case 2*: The matrix dyadic logical operators for implication L and disjunction D defined for Set 1 and Set 2.

Set 1: $L = \begin{bmatrix} 1 & 0 & 1 & 1 \\ 0 & 1 & 0 & 0 \end{bmatrix}, \ D = \begin{bmatrix} 1 & 1 & 1 & 0 \\ 0 & 0 & 0 & 1 \end{bmatrix}$ .

Set 2: $L = \frac{1}{\sqrt{2}}\begin{bmatrix} 2 & 0 & 0 & 0 \\ 1 & 1 & -1 & 1 \end{bmatrix}, \ D = \frac{1}{\sqrt{2}}\begin{bmatrix} 2 & 0 & 0 & 0 \\ 1 & 1 & 1 & -1 \end{bmatrix}..$

## 3. The Square Roots of Not.

Note that the arithmetic representation we showed earlier, based on the pair $\{1, -1\}$, allows us to establish that $\text{Not} = -1$. From here come, for this scalar representation, the two complex roots of negation:

$$\left(\sqrt{\text{Not}}\right)_1 = i \quad \text{and} \quad \left(\sqrt{\text{Not}}\right)_2 = -i \quad \text{with } i = \sqrt{-1} \ .$$

These two roots imply that $\left(\sqrt{\text{Not}}\right)_1 . \left(\sqrt{\text{Not}}\right)_2 = 1$, that for $\tau = \{1, -1\}$ gives the identity $\text{Id} = 1$.

We will now extend the search for the square roots of negation to the matrix domain. As noted in the Introduction, this matrix square root of Not is an important concept in the analysis of the novel potentialities of quantum computing [6]. In the framework of quantum computing, this operator is a complex square matrix of dimension 2, and operates on qubits, 2D vectors. Here, we will extend the definition of the square roots of negation to truth vectors of arbitrary dimension, as vectors $s, n \in \mathbb{R}^{Q\times 1}$ used in the previous section.

To simplify the calculations, we will start by analyzing the situation in which both vectors are orthonormal [5]. Our problem is the following: given the operator

$$N = ns^T + sn^T,$$

find a root $\left(\sqrt{N}\right)_1$. We begin by establishing, as a test solution, the following two equations:

$$\left(\sqrt{N}\right)_1 s = \alpha s + \beta n \; ; \; \left(\sqrt{N}\right)_1 n = \alpha' s + \beta' n \; . \qquad (10)$$

Since $Ns = n$ and $Nn = s$, we obtain

$$\left(\sqrt{N}\right)_1 \left[\left(\sqrt{N}\right)_1 s\right] = n \; ; \; \left(\sqrt{N}\right)_1 \left[\left(\sqrt{N}\right)_1 n\right] = s \; . \qquad (11)$$

Substituting equations (10) in (11), we obtain the following relationships between the coefficients:

(A) $\alpha^2 + \beta\alpha' = 0$ and $\alpha\beta + \beta\beta' = 1$ .

(B) $\alpha'\alpha + \beta'\alpha' = 1$ and $\alpha'\beta + \beta'^2 = 0$ .

Analysis of equations (A) and (B) produces the following important results:

R1. Symmetry. $\alpha' = \beta$ and $\beta' = \alpha$ .

R2. $\alpha^2+\beta^2=0$ and $2\alpha\beta=1$ .

R3. $\alpha+\beta=1$ .

Note that R3 is a nice corollary of R2:

$$(\alpha+\beta)^2=(\alpha^2+\beta^2)+2\alpha\beta=1 \ .$$

Selecting the positive root, we obtain R3, $\alpha+\beta=1$.

The first equation for R2 indicates that $\alpha$ and $\beta$ are complex numbers. If we write down $\alpha=u+iv$ and $\beta=u-iv$, it can be shown from R2 y R3 that

$$\alpha=\tfrac{1}{2}(1+i) \quad \text{and} \quad \alpha=\tfrac{1}{2}(1-i) \ .$$

Let us now construct an equation in which $\left(\sqrt{N}\right)_1$ is an unknown. For this we use equation (10), already assuming the symmetry of the numerical coefficients, and we obtain

$$\left(\sqrt{N}\right)_1[s \ n]=[(\alpha s+\beta n) \quad (\beta s+\alpha n)] \tag{12}$$

and since, by the orthonormality of s and n, the pseudo-inverse is $[s \ n]^{+}=[s \ n]^{T}$, finally we have that the solution of this equation is

$$\left(\sqrt{N}\right)_1=[(\alpha s+\beta n) \quad (\beta s+\alpha n)][s \ n]^{T} \ . \tag{13}$$

Developing this matrix equation and using the complex values of $\alpha$ and $\beta$, we obtain this beautiful equation:

$$\left(\sqrt{N}\right)_1=\tfrac{1}{2}(1+i)I+\tfrac{1}{2}(1-i)N \ , \tag{14}$$

with I and N being, respectively, the matrix identity and matrix negation shown in the previous section. An analogous calculation gives us the second square root of N, whose structure is

$$\left(\sqrt{N}\right)_2 = \tfrac{1}{2}(1+i)\,N + \tfrac{1}{2}(1-i)\,I . \tag{15}$$

These two roots are complex conjugates and its product gives the logical identity I : $\left[\left(\sqrt{N}\right)_2\right]^{*} = \left(\sqrt{N}\right)_1$ and $\left(\sqrt{N}\right)_1 . \left(\sqrt{N}\right)_2 = I$.

To finish this section, let us point out that the same calculation can be made for a pair of normal and linearly independent vectors s and n. In this case, the following identity $\overline{I}$ and negation $\overline{N}$ matrices are obtained from equation (7) :

$$\overline{I} = s y^T + n z^T \quad \text{and} \quad \overline{N} = n y^T + s z^T \ .$$

In this case, the square roots of $\overline{N}$ are similar to those given in equations (14) and (15), changing I by $\overline{I}$ and N by $\overline{N}$.

In the following explorations, to simplify the arguments, we will use the following notation:

$$A = \left(\sqrt{N}\right)_1 \quad , \quad B = \left(\sqrt{N}\right)_2 . \tag{16}$$

With this simple notation, we can summarize some interesting properties of the roots of the matrix N:

$$A^2 = N \ ; \ B^2 = N \ ; \ B = A^{*} \ ; \ AB = BA = I \ ; \ B = NA \ ; \ B = NA .$$

To finish this section, we present a 4D numerical example. Let us consider the following two orthonormal truth vectors: $s = \tfrac{1}{2}[1 \quad 1 \quad 1 \quad 1]^T$ and $n = \tfrac{1}{2}[1 \quad -1 \quad -1 \quad 1]^T$. The associated identity and negation matrices I and N are

$$I = \frac{1}{2}\begin{bmatrix} 1 & 0 & 0 & 1 \\ 0 & 1 & 1 & 0 \\ 0 & 1 & 1 & 0 \\ 1 & 0 & 0 & 1 \end{bmatrix} \quad ; \quad N = \frac{1}{2}\begin{bmatrix} 1 & 0 & 0 & 1 \\ 0 & -1 & -1 & 0 \\ 0 & -1 & -1 & 0 \\ 1 & 0 & 0 & 1 \end{bmatrix} ,$$

and the two square roots of N are

$$A = \frac{1}{2}\begin{bmatrix} 1 & 0 & 0 & 1 \\ 0 & i & i & 0 \\ 0 & i & i & 0 \\ 1 & 0 & 0 & 1 \end{bmatrix} \quad ; \quad B = \frac{1}{2}\begin{bmatrix} 1 & 0 & 0 & 1 \\ 0 & -i & -i & 0 \\ 0 & -i & -i & 0 \\ 1 & 0 & 0 & 1 \end{bmatrix}$$

We note in passing that the eigenvalues of A are (1,0,i,0) and those of B are (1,0,-i,0).

## 4. Explorations.

This is the heuristic part of this article. Here we will explore two topics: 1) the possibility that when faced with a hidden logic gate, the preprocessing of a single input using the square roots of Not allows a full identification of which is the hidden logic gate, and 2) the possibility of obtaining fully matrix versions of the Euler equation for the complex exponential using, instead of the imaginary unit I, one of the square roots of Not.

### 4.1. Global information from partial data.

This exploration is inspired by a surprising result obtained by Deutsch et al [6] in the domain of quantum computing. We begin by stating the simplest form of the Deutsch's problem.

Let us imagine the following situation: a logic gate is inside a black box that allows only an exploratory action that consists of introducing a single input and collecting the output. So, is it possible under these conditions to find out which is the logic gate locked in the black box? We can state the problem in the following way. Table 1 shows that if before a hidden monadic logical operation, we introduce the truth value t as input, two options arise: the output is t, and it is generated by Id and Cid, or the output is f and it is generated by Not or Cnot. This isolated input does not allow distinguishing constant operations from operations whose output depends on the input. Now, the remarkable result of Deursch et al. [3] obtained in the context of quantum computing (shows that if instead of using t we use qubit $|1\rangle$ (a 2D column vector) as the truth value and preprocess it by $\sqrt{N}$, it becomes possible to diagnose whether our hidden operation is a function variable or a constant function. This result is usually known as "Deutsch's algorithm" ([2], p.145).

Here, our approach is not quantum, but relies on the operators shown above, and on actual Q-dimensional truth values. The result of Deutsch et al, previously mentioned, suggests that we analyze for the monadic operators the effect of the prefilter of the input s by a $\sqrt{N}$, and for the dyadic operators corresponding to Table 2, study the effect of prefiltering the only input $(s \otimes s)$ by means of a $\left(\sqrt{N} \otimes \sqrt{N}\right)$ operator. We believe that the results that we will show below are suggestive. It must be emphasized that in this exploration, we rely only on the basic laws of linear algebra and the projections of outputs on hyperplanes, rather than the physical conditions of quantum operators.

We start by studying monadic operators.

a1). *Monadic operators.* We are going to use only s as input, but we premultiply the input by the root A. The results are the following:

$$KAs = \tfrac{1}{2}(1+i)s + \tfrac{1}{2}(1-i)s \; ; \; MAs = \tfrac{1}{2}(1+i)n + \tfrac{1}{2}(1-i)n \; ;$$

$$IAs = \tfrac{1}{2}(1+i)s + \tfrac{1}{2}(1-i)n \; ; \; NAs = \tfrac{1}{2}(1+i)n + \tfrac{1}{2}(1-i)s \, .$$

The important point is that only one input displays the whole outputs of the operator (normally univocally with each one of the potential inputs).

Separating real and imaginary terms, we get

$$KAs = s \; ; \; MAs = n \; ;$$
$$IAs = \tfrac{1}{2}(s+n) + i\tfrac{1}{2}(s-n) \; ; \; NAs = \tfrac{1}{2}(n+s) - i\tfrac{1}{2}(s-n) \, .$$

This distinction of the operators I and N is not possible in quantum computing due to the disappearance of the scalar signs. Since we are not in the quantum domain, the signs of the scalar coefficients remain, and the consequence is that $IAs$ and $NAs$ have the same real part, but the imaginary terms are opposite vectors. *The remarkable conclusion is that if we didn't know the operator, the sole input s premultiplied by the root A provides us with a complete diagnostic of the operator involved.* This mathematical fact represent an extreme solution of the problem of uncover hidden monadic operators, that in this geometrical, non-quantum, approach allow us to detect not only constant from "balanced" functions, but to obtain the complete diagnosis of the four monadic functions.

a2) *Dyadic operators.* Now we can explore what happen with dyadic operations if we use as sole input the pair $(s \otimes s)$. We begin by premultiplying this input by the tensor product $(A \otimes A)$. The outputs produced are the following:

$$C(A \otimes A)(s \otimes s) = \tfrac{1}{2}is + \tfrac{1}{2}n + \tfrac{1}{2}n - \tfrac{1}{2}in =$$
$$n + \tfrac{1}{2}i(s-n);$$

$$D(A \otimes A)(s \otimes s) = \tfrac{1}{2}is + \tfrac{1}{2}s + \tfrac{1}{2}s - \tfrac{1}{2}in$$
$$= s + \tfrac{1}{2}i(s-n);$$

;

$$L(A \otimes A)(s \otimes s) = \tfrac{1}{2}is + \tfrac{1}{2}n + \tfrac{1}{2}s - \tfrac{1}{2}is$$
$$= \tfrac{1}{2}(n+s);$$

;

$$E(A \otimes A)(s \otimes s) = \tfrac{1}{2}is + \tfrac{1}{2}n + \tfrac{1}{2}n - \tfrac{1}{2}is$$
$$= n;$$

$$X(A \otimes A)(s \otimes s) = \tfrac{1}{2}in + \tfrac{1}{2}s + \tfrac{1}{2}s - \tfrac{1}{2}in$$
$$= s.$$

We see that this preprocessing of the input $(s \otimes s)$ by $(A \otimes A)$, similar to what happened with the monadic functions, displays all the outputs of the dyadic functions C, D, L, E and X, and the projections of these unfolded outputs in the real an the imaginary Q-dimensional hyperplanes are different for these different functions. This remarkable property is also valid for Nand and Nor:

$$NAND = NC = s - \tfrac{1}{2}i(s-n) \quad \text{and} \quad NOR = NC = n - \tfrac{1}{2}i(s-n).$$

*Consequently, these seven operations produce seven different projections in the real and imaginary hyperplanes.* It is important to remark that this fact is not valid for the whole set of 16 dyadic functions due to redundancy of symmetries in the first and last terms and in the central terms. Finally, we point out that this distinction of these seven operators is not possible in quantum computing due to the disappearance of the signs of the scalars.

### 4.2. Fully matrix Euler expansions.

To support this exploration, we are going to present some classic results. First, we start with the expansion of the exponential:

$$e^{x} = 1 + x + \frac{x^2}{2!} + \frac{x^3}{3!} + \frac{x^4}{4!} + \cdots \qquad (17)$$

As is well known, the series (17) converges for all $x \in \mathbb{R}$. Let us now look at this expansion for the complex exponent:

$$e^{i\varphi} = \left(1 - \frac{\varphi^2}{2!} + \frac{\varphi^4}{4!} - \cdots\right) + i\left(\varphi - \frac{\varphi^3}{3!} + \frac{\varphi^5}{5!} - \cdots\right) \qquad (18)$$

where the series in parentheses are the classic expansions of $\cos(\varphi)$ and $\sin(\varphi)$. Hence the famous Euler equation

$$e^{i\varphi} = \cos(\varphi) + i\sin(\varphi)\ . \qquad (19)$$

De Moivre's formula can be seen as a corollary of this equation:

$$e^{i\varphi n} = \left(\cos(\varphi) + i\sin(\varphi)\right)^{n} = \cos(\varphi n) + i\sin(\varphi n)\ .$$

Let us now look at the Eulerian formulas for circular functions:

$$\cos^2(\varphi) + \sin^2(\varphi) = e^{i\varphi}e^{-i\varphi} = 1\,;$$

$$\sin(\varphi) = -\mathrm{i}\tfrac{1}{2}\left(e^{\mathrm{i}\varphi} - e^{-\mathrm{i}\varphi}\right) \;\; ; \quad \cos(\varphi) = \tfrac{1}{2}\left(e^{\mathrm{i}\varphi} + e^{-\mathrm{i}\varphi}\right);$$

$$\sin(\alpha+\beta) = -\mathrm{i}\tfrac{1}{2}\left(e^{\mathrm{i}\alpha}e^{\mathrm{i}\beta} - e^{-\mathrm{i}\alpha}e^{-\mathrm{i}\beta}\right) = \sin(\alpha)\cos(\beta) + \sin(\beta)\cos(\alpha) \;\; ;$$

$$\cos(\alpha+\beta) = \tfrac{1}{2}\left(e^{\mathrm{i}\alpha}e^{\mathrm{i}\beta} + e^{-\mathrm{i}\alpha}e^{-\mathrm{i}\beta}\right) = \cos(\alpha)\cos(\beta) - \sin(\alpha)\cos(\beta) \;\; .$$

Finally, let us point out that from (19), doing $\varphi = \pi$ the Great Euler Equation is obtained, which appears on the podium of the most beautiful equations of Mathematics:

$$e^{\mathrm{i}\pi} + 1 = 0 \;\; .$$

From these classical results, our exploration consists of developing matrix exponentials in which the imaginary variable i is replaced by one of the square roots of Not and analyzing whether there are fully matrix (f-m)versions for the preceding expressions.

Let us start by showing the classic expression of the matrix exponential for a $G \in \mathbb{C}^{Q\times Q}$ matrix as a convergent power series:

$$e^{G} = I + G + \frac{G^2}{2!} + \frac{G^3}{3!} + \frac{G^4}{4!} + \cdots \quad . \tag{20}$$

Let us now look up a f-m version of the complex exponential $e^{\mathrm{i}\varphi}$ by doing $G = AX$ where $A = \left(\sqrt{N}\right)_1$ represents a matrix version of $\mathrm{i} = \sqrt{-1}$ . The strong hypothesis that we will assume to support the following development is that A and X commute, therefore $(AX)^n = A^n X^n$ .

Under these conditions, the result is

$$e^{AX} = I + AX + \frac{A^2X^2}{2!} + \frac{A^3X^3}{3!} + \frac{A^4X^4}{4!} + \frac{A^5X^5}{5!} + \cdots \quad . \tag{21}$$

But $A^2 = N$, $A^3 = AN$, $A^4 = N^2 = I$, $A^5 = AI = A$, etc. Consequently,

$$e^{AX} = \left( I + N\frac{X^2}{2!} + \frac{X^4}{4!} + N\frac{X^6}{6!} + \cdots \right) + A\left( X + N\frac{X^3}{3!} + \frac{X^5}{5!} + N\frac{X^7}{7!} + \cdots \right). \tag{22}$$

Imitating equation (19) we can put

$$e^{AX} = C(X) + AS(X), \tag{23}$$

with C (X) and S (X) are given by the series

$$C(X) = I + N\frac{X^2}{2!} + \frac{X^4}{4!} + N\frac{X^6}{6!} + \cdots \quad ,$$

$$S(X) = X + N\frac{X^3}{3!} + \frac{X^5}{5!} + N\frac{X^7}{7!} + \cdots \quad .$$

It should be noted that these equations (23) are not the classical matrix equations of $\cos(X)$ and $\sin(X)$ (described, for example, in [1], p. 245, problem 9.7). The difference is that the $-1$ of the $\cos(X)$ and $\sin(X)$ matrices that we have just mentioned, is replaced by the negation matrix N in C (X) and S (X).

Let us now assume that matrices A, X and Z commute. Under this conditions we have

$$e^{AXZ} = C(XZ) + AS(XZ). \tag{24}$$

But note that taking into account eq. (23), this is equivalent to putting

$$e^{AXZ} = C(XZ) + AS(XZ) \equiv \left[ C(X) + AS(X) \right]^{Z} . \tag{25}$$

So we have a f-m matrix version of De Moivre's equation, for $Z \in \mathbb{R}^{Q\times Q}$. We comment in passing that eq. (25) suggests the interesting problem of the conditions of validity of the formula $(e^{G})^{Z} = e^{GZ}$ (in our case, $e^{G}$ = equation (23)).

Remark that if in equation (25) $Z = O$ we get $C(O) = I$ and $S(O) = O$ (O is the matrix zero, but I is the logical identity).

The next step in this game with linear algebra is to try to obtain f-m versions of the Eulerian circular functions. This is going to require a proposal that may be extremely trivial, but it is still a way to achieve matrix versions for circular functions.
We will first look for the f-m version of $\cos^2(\varphi) + \sin^2(\varphi) = e^{i\varphi}e^{-i\varphi} = 1$.

$$e^{AX}e^{-AX} = \left[C(X) + AS(X)\right]\left[C(X) - AS(X)\right] .$$

Developing we get

$$C^2(X) - NS^2(X) = I$$

Let us follow with $\cos(\varphi) = \frac{1}{2}\left(e^{i\varphi} + e^{-i\varphi}\right)$. The corresponding matrix version is

$$C(X) = \frac{1}{2}\left(e^{AX} + e^{-AX}\right).$$

Taking into account that the matrix version of $-i$ is B, we found that $\sin(\varphi) = -i\frac{1}{2}\left(e^{i\varphi} - e^{-i\varphi}\right)$ corresponds with matrix

$$S(X) = \frac{1}{2}B\left(e^{AX} - e^{-AX}\right) .$$

The expression $\cos(\alpha+\beta)=\frac{1}{2}\left(e^{i\alpha}e^{i\beta}+e^{-i\alpha}e^{-i\beta}\right)=\cos(\alpha)\cos(\beta)-\sin(\alpha)\sin(\beta)$ can be translated to a matrix format as follows using X and the third commutative matriz Z:

$$C(X+Z)=\frac{1}{2}\left(e^{AX}e^{AZ}+e^{-AX}e^{-AZ}\right) .$$

The commutation of the matrix exponents allows to put $e^{X}e^{Z}=e^{X+Z}$, and similarly for the negative exponents. After developing the previous expression and simplifying, we obtain our final result:

$$C(X+Z)=C(X)C(Z)+NS(X)S(Z).$$

The expression $\sin(\alpha+\beta)=-i\frac{1}{2}\left(e^{i\alpha}e^{i\beta}-e^{-i\alpha}e^{-i\beta}\right)=\sin(\alpha)\cos(\beta)+\sin(\beta)\cos(\alpha)$ inspire the matrix format

$$S(X+Z)=\frac{1}{2}B\left(e^{AX}e^{AZ}-e^{-AX}e^{-AZ}\right)$$

and after operating and simplifying, we obtain

$$S(X+Z)=S(X)C(Z)+S(Z)C(X) .$$

The final point is to look for a f-m version for the Great Euler Equation

$$e^{i\pi}+1=0.$$

This involves discovering a matrix $\Pi$ such that $e^{A\Pi}=-I+O$. A trivial choice is to define $\Pi=Bi\pi$. Is there any other matrix that achieves this?

## 5. Conclusions.

It has been a magnificent fact, repeated throughout history, that the study of Nature poses new and interesting challenges to mathematical research. In the case we are dealing with in this article, the immersion of logical operations in linear algebra structures is a territory that offers wide avenues for future exploration. The study of the properties of the square roots of the matrix operator N, which represents logical negation, is a specific example that illustrates one of the various perspectives that open up for logic due to the transformation of the formal operations of classical algebraic logic in a theory of operators, based on matrices and vectors.

**Acknowledgments.** The author thanks to the Agencia Nacional de la Investigación y la Innovación (ANII-Uruguay) for partial financial support.